\documentclass[aps,floatfix,twocolumn,tightenlines,amsmath,amssymb,nofootinbib]{revtex4-1}

\usepackage[utf8]{inputenc}
\usepackage[T1]{fontenc}
\usepackage{graphicx}
\usepackage{amssymb}
\usepackage{color}
\usepackage{amsthm}
\usepackage{psfrag}
\usepackage{ifsym}
\usepackage{epstopdf}
\usepackage{dsfont}
\usepackage{amsfonts}
\usepackage{multirow}
\usepackage{bm}
\usepackage{bbm}
\usepackage{mathtools}

\newcommand{\expec}[1]{\left\langle #1 \right\rangle}
\newcommand{\bra}[1] {\langle #1 |}
\newcommand{\ket}[1] {| #1 \rangle}

\newcommand{\one}{\leavevmode\hbox{\small1\normalsize\kern-.33em1}}

\newcommand{\figref}[1]{Fig.~\ref{#1}}

\newcommand\id{\leavevmode\hbox{\small1\kern-3.3pt\normalsize1}}

\def\vv{\mathbf{v}}
\def\d{\mathbf{d}}
\def\e{\mathbf{e}}
\def\q{\mathbf{q}}
\def\p{\mathbf{p}}
\def\RR{\mathbbm{R}}
\def\1{\mathbf{1}}
\def\0{\mathbf{0}}

\def\minimize{\textrm{minimize}}
\def\st{\textrm{subject to }}

\begin{document}
\title{Experimental Test of Nonlocal Causality}

\author{M. Ringbauer$^{1,2}$, C. Giarmatzi$^{1,2}$, R. Chaves$^{3,4}$, F. Costa$^{1}$, A. G. White$^{1,2}$ \& A. Fedrizzi$^{1,2,5}$}
\affiliation{$^{1}$Centre for Engineered Quantum Systems, $^{2}$Centre for Quantum Computer and Communication Technology, School of Mathematics and Physics, University of Queensland, Brisbane, QLD 4072, Australia,\\$^{3}$Institute for Physics \& FDM, University of Freiburg, 79104 Freiburg, Germany, $^{4}$Institute for Theoretical Physics, University of Cologne, 50937 Cologne, Germany,\\$^{5}$School of Engineering and Physical Sciences, SUPA, Heriot-Watt University, Edinburgh EH14 4AS, UK}

\begin{abstract}
\noindent Explaining observations in terms of causes and effects is central to all of empirical science. Correlations between entangled quantum particles, however, seem to defy such an explanation. To recover a causal picture in this case, some of the fundamental assumptions of causal explanations have to give way. 
Here we consider a broad class of models where one of these assumptions, Bell's local causality, is relaxed by allowing a direct influence from one measurement outcome to the other. 
We use interventional and observational data from a photonic experiment to bound the strength of this causal influence in a two-party Bell scenario and test a novel Bell-type inequality for the considered models. Our results demonstrate the incompatibility of quantum mechanics with an important class of nonlocal causal models, which includes Bell's original model as a special case. Recovering a classical causal picture of quantum correlations thus requires an even more counter-intuitive modification of our classical notion of cause and effect.
\end{abstract}

\maketitle
\begin{figure}[!h]
\includegraphics[width=1\columnwidth]{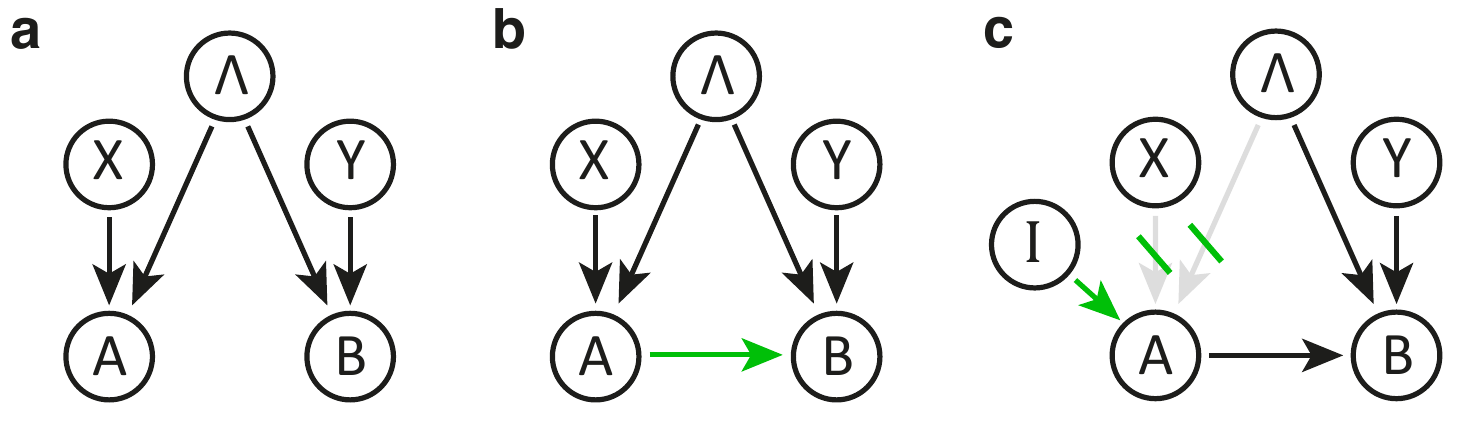}
\caption{\textbf{Causal structures for a Bell scenario.} \textbf{(a)} Bell's original local hidden variable model, where $X$ ($Y$) is Alice's (Bob's) measurement setting and $A$ ($B$) is the corresponding measurement outcome. $\Lambda$ denotes the local hidden variable. \textbf{(b)} A relaxation of local causality, where $A$ may have direct causal influence on $B$. Bell's model in (a) is the limiting case where the green arrow from $A$ to $B$ vanishes. \textbf{(c)} An intervention ($I$) on $A$ forces the variable to take a specific value and breaks all incoming arrows.}
\label{fig:models}
\end{figure}

\noindent 
Four decades after Freedman and Clauser~\cite{Freedman1972} performed the first Bell-inequality test \cite{Bell1964}, a series of loophole-free experiments~\cite{hensen2015experimental,Shalm2015,Giustina2015} have now conclusively shown that the predictions of quantum mechanics are at odds with the world view of local realism. Scientific realism posits that physical systems have real, objective properties---independent of whether we observe them or not---that determine the outcomes of measurements performed on the system. The idea of locality, or more precisely \emph{local causality} is that causal influences cannot propagate faster than the speed of light. Based on local causality, and the assumption that measurement settings can be chosen freely, Bell derived an inequality that must be respected by any set of correlations that can be explained in terms of, possibly hidden, common causes, cf.\ Fig.~\ref{fig:models}a, but is violated by observed quantum correlations.
Consequently, a new area of research has emerged, exploring to what extent the various underlying assumptions have to be relaxed in order to recover a causal explanation of quantum correlations~ \cite{Brans1988,Branciard2008,Hall2010,Hall2011,Barrett2010,Chaves2015b,Spekkens2015,Aktas2015,Putz2015}.

An excellent platform for this research program, and a natural framework for Bell's theorem, is the theory of causal modelling~\cite{Spekkens2015,Chaves2015b}, which aims to explain correlations in terms of cause-and-effect relations between events~\cite{Pearlbook,Spirtesbook}. Discovering these relations from empirical data is difficult in general~\cite{Geiger1999,tian2002testable,Chaves2014b,mooij2014distinguishing}, however, within classical physics such an explanation should always exist, since the properties of a classical system, even if not measured, can always be assumed to have well-defined values. Such causal reasoning is at the heart of empirical science and builds upon the most fundamental understanding of causality, that if a variable acts as the cause for another one, actively intervening on the first should cause changes in the second. More recently, causal modelling has attracted considerable interest in foundational physics, in particular for the study of stronger-than-classical correlations~\cite{Spekkens2015,Chaves2015a,Fritz2012,Fritz2014,cavalcanti2014modifications,pienaar2014graph,Leifer2013,Henson2015}, dynamical causal order~\cite{Oreshkov2015}, and indefinite causal structures~\cite{Brukner2014,Oreshkov2015} and their role as computational resource~\cite{Chiribella2013, chiribella_perfect_2012, araujo2014computational, Procopio2014}.

Phrasing Bell's theorem in the language of causal models provides a clear picture of the underlying assumptions and allows for a unified and quantitative approach to relaxations of these assumptions~\cite{Spekkens2015,Chaves2015b}. For example, causal models can in principle reproduce quantum correlations when relaxing Bell's local causality assumption, which is commonly referred to as \emph{quantum nonlocality}. Here we test models which allow for a causal influence from one measurement outcome to the other, cf. Fig.~\ref{fig:models}b. First we consider the simplest and most well-studied example of such correlations, the Clauser-Horne-Shimony-Holt (CHSH) scenario~\cite{Clauser1969}, where two parties, Alice and Bob, can each measure one of two dichotomic observables. Using controlled interventions we find the potential causal influence insufficiently strong to explain the observed CHSH violation. In the second experiment we go beyond the simple CHSH scenario and violate a novel Bell-type inequality which involves three measurement settings for each party and is satisfied even for arbitrarily strong causal influences from one outcome to the other~\cite{Chaves2015b}. In contrast to the interventional method, which requires detailed knowledge of the physical system under consideration, the latter method is device-independent. Our results highlight the incompatibility of quantum correlations, not just with Bell's local causal model, but even with nonlocal causal models where one measurement outcome may have a direct causal influence on the other.

\textbf{Causal modeling ---}
A causal structure underlying $n$ jointly distributed discrete random variables $(X_1, \dots, X_n)$ is represented by a directed acyclic graph (DAG), with the nodes (circles in Fig.~\ref{fig:models}) representing variables and the directed edges (arrows in Fig.~\ref{fig:models}) representing causal relations~\cite{Pearlbook}. Bell's theorem, where two observers, Alice and Bob, perform local measurements on one half of a shared quantum state, can be conveniently formulated in this language. Figure~\ref{fig:models}a shows the corresponding causal graph, based on Bell's assumptions of measurement independence and local causality. 
Measurement independence states that the measurement choices of Alice and Bob, $X$ and $Y$ respectively, are independent of how the system has been prepared, i.e.\ there is no causal link from the hidden variable $\Lambda$ to $X$ or $Y$ and thus $p(x,y,\lambda)=p(x,y)p(\lambda)$~\footnote{We adopt the usual convention that uppercase letters label random variables while their values are denoted in lower case.}. Local causality implies that the probability of Alice's (Bob's) outcome $A$ ($B$) is fully specified by $\Lambda$ and by the measurement choice $X$ ($Y$), i.e.\ $p(a\vert x,y,b,\lambda)=p(a\vert x,\lambda)$ ($p(b\vert x,y,a,\lambda)=p(b\vert y,\lambda)$). The latter assumption is reflected in the causal graph, Fig.~\ref{fig:models}a, by what we call \emph{causal parameter independence}---there is no direct causal influence from the measurement setting $Y$ ($X$) to the other party's outcome $A$ ($B$)---and \emph{causal outcome independence}, stating that there is no direct causal influence from one outcome to the other~\footnote{Note that these definitions differ slightly from their statistical variants, see Supplementary Information for details, see Sec.~\ref{Sec:Supp1}.}.

The causal models compatible with these assumptions are of the form of Bell's well-known local hidden variable model: $p(a,b \vert x,y) = \sum_{\lambda} p(a\vert x, \lambda) p(b\vert y, \lambda) p(\lambda)$ .The constraints on the observable probabilities $p(a,b \vert x,y)$ dictated by such a causal model are known as Bell inequalities. In the simplest possible Bell scenario, where each of the parties measure one of two observables ($x,y=0,1$) obtaining one of two possible outcomes ($a,b=0,1$), any correlations compatible with Bell's causal model must respect the CHSH-inequality~\cite{Clauser1969}
\begin{align}
S_2 = \expec{A_0B_0}+\expec{A_0B_1}+\expec{A_1B_0}-\expec{A_1B_1} \leq 2 ,
\label{CHSH}
\end{align}
where $\expec{A_xB_y}=\sum_{a,b=0,1}(-1)^{a+b}p(a,b\vert x,y)$ is the joint expectation value of $A_x$ and $B_y$. The first loophole-free Bell experiments~\cite{hensen2015experimental,Shalm2015,Giustina2015} now conclusively show that quantum mechanics allows for correlations that violate this inequality, therefore witnessing its incompatibility with causal models satisfying local causality and measurement independence.

In order to retain a classical causal explanation of the Bell scenario, some of these causal assumptions have to be relaxed~\cite{Branciard2008,Hall2010,Hall2011,Barrett2010,Spekkens2015,Chaves2015b,Aktas2015,Putz2015,WisCav16}. 
We focus on the class of models which do not assume causal outcome independence, such that Alice's measurement outcomes may have a direct causal influence on Bob's outcomes (or vice-versa), while satisfying causal parameter independence, see Fig.~\ref{fig:models}b. Since the causal model is formulated without any reference to a space-time structure, this influence may be sub- or superluminal, instantaneous, or even to the past, as long as it does not create any causal loop. In particular, it is consistent with a recent no-go theorem stating that quantum correlations cannot be explained by any finite-speed influence~\cite{Bancal2012a}. The probability distributions compatible with this causal structure can be decomposed as 
\begin{equation}
\label{eq:JointProb}
p(a,b \vert x,y) = \sum_{\lambda} p(a\vert x, \lambda) p(b\vert y,a, \lambda) p(\lambda) .
\end{equation}

The first experimental method we use to test this model relies on \emph{interventions}, a core tool in causal discovery allowing for the identification and quantification of causal influences~\cite{Pearlbook,janzing2013quantifying,Ried2015,Chaves2015b}. Formally, an intervention is the act of locally forcing a variable $X_i$ to take on some value $x^{\prime}_i$, denoted $\mathrm{do}(x^{\prime}_i)$. This removes all incoming arrows on $X_i$, while keeping the causal dependencies between all other variables unperturbed, see $A$ in Fig.~\ref{fig:models}c.

In the CHSH-scenario, passive observations alone are not enough to determine whether correlations between $A$ and $B$ are due to direct causation or a common cause $\Lambda$. An intervention on variable $A$, however, would break the link between $A$ and the (hypothetical) variable $\Lambda$. All remaining correlations between $A$ and $B$ must thus stem from direct causation. Indeed, the maximal shift in the probability distribution of $B$ upon intervention on $A$ even allows quantifying the strength of this causal link~\cite{Chaves2015b}. To achieve this we use the so-called \emph{average causal effect}~\cite{Pearlbook,janzing2013quantifying}, 
\begin{equation}
\label{eq:ACE}
\mathrm{ACE}_{A \rightarrow B}= \sup_{b,y,a,a^{\prime}}  \vert p(b\vert \mathrm{do}(a, y)-p(b\vert \mathrm{do}(a^{\prime}), y)\vert ,
\end{equation}
which is a variant of the measure $\mathcal{C}_{A \rightarrow B}$ used in Ref.~\cite{Chaves2015b}. In contrast to this measure, however, $\mathrm{ACE}_{A \rightarrow B}$ does not require knowledge of the hidden variable and is thus experimentally accessible. As we prove in detail in Sec.~\ref{Sec:Supp1}, the average causal effect satisfies the same relation as $\mathcal{C}_{A \rightarrow B}$ in Ref.~\cite{Chaves2015b}, namely,
\begin{equation}
\min \mathrm{ACE}_{A \rightarrow B}=\max \left[ 0,(S_2-2)/2 \right] ,
\label{eq:ACE_CHSH}
\end{equation}
where the maximum is taken over all eight symmetries of the CHSH quantity under relabelling of inputs, outputs, and parties~\cite{Clauser1969}. In other words the minimal average causal effect required for a causal explanation of a set of quantum correlations is directly proportional to the CHSH violation achieved by the correlations in question. 

\begin{figure}[t!]
  \begin{center}
\includegraphics[width=\columnwidth]{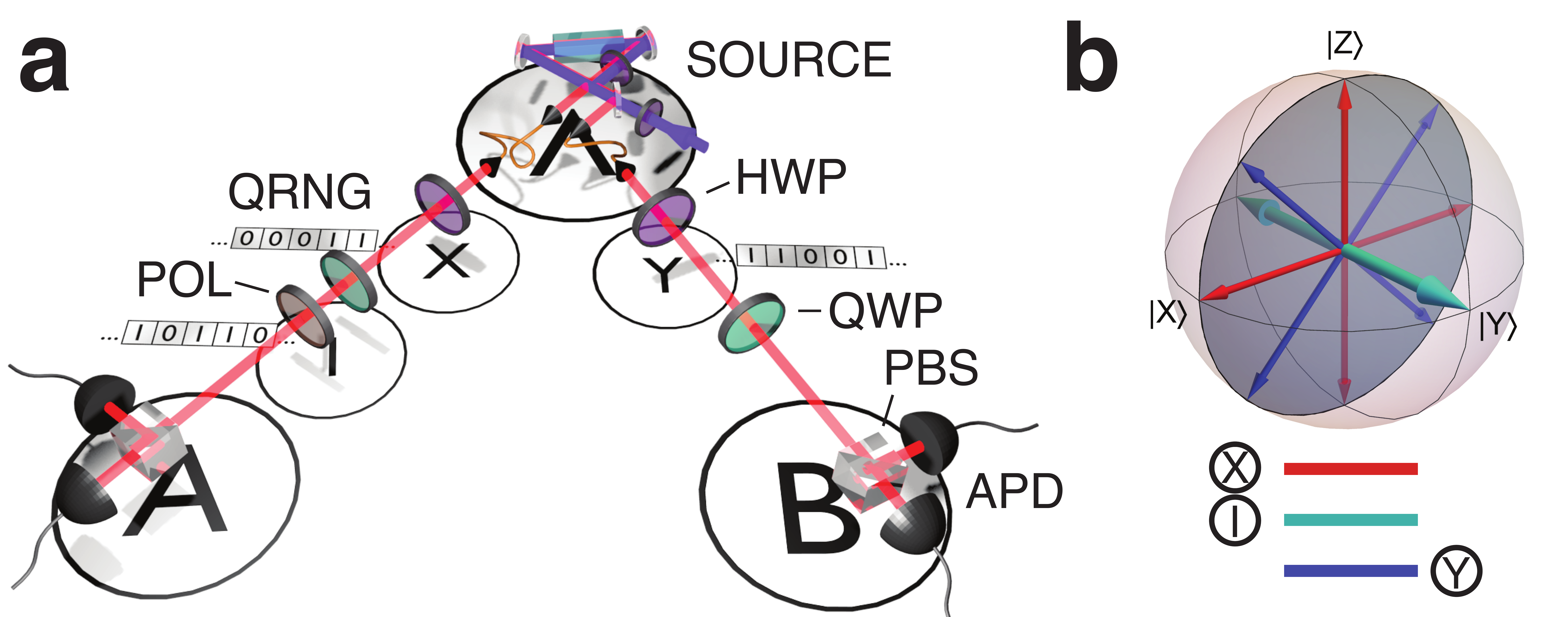}
  \end{center}
\vspace{-5mm}
\caption{\textbf{(a) The experimental setup}. Pairs of photons are generated via spontaneous parametric downconversion in a periodically poled KTP (ppKTP) crystal, using the Sagnac design of Ref.~\cite{Fedrizzi2007}. The degree of polarization entanglement between the two photons can be continuously varied by changing the polarization-angle $\gamma$ of the pump laser. Alice and Bob perform measurements in the equatorial plane of the Bloch sphere using a half-wave plate (HWP) and a polarizing beam splitter (PBS). Additional quarter-wave plates (QWP) can be used for quantum state tomography of the initial entangled state. In the interventionist experiment an additional combination of QWP and polarizer (POL) are used between Alice's basis choice and her measurement. Causal variables are indicated using the notation of Fig.~\ref{fig:models}a. \textbf{(b)} Alice's (red) and Bob's (blue) measurement bases and the intervention direction (cyan) on the Bloch-sphere.}
  \label{fig:Setup}
\end{figure}

\textbf{Interventional method ---}
We experimentally implemented an intervention on a CHSH-Bell test using pairs of polarization-entangled photons, generated in the state $\cos\gamma \ket{HV} + \sin\gamma \ket{VH}$, see Fig.~\ref{fig:Setup}a. Here $H$ and $V$ correspond to horizontal and vertical polarizations, respectively and $\gamma$ is the polarization angle of the pump beam, which continuously controls the degree of entanglement, as measured by the concurrence $\mathcal{C}=|\sin2\gamma|$~\cite{Hill1997}.

Alice and Bob perform a standard CHSH-inequality test with two settings and two outcomes each. The measurements are chosen in the equatorial (linear-polarization) plane of the Bloch sphere, see Fig.~\ref{fig:Setup}b. In order to test the (directional) link $A\to B$, Bob has been located in the causal future of Alice using a $2$~m fibre delay before Bob's measurement device. An intervention on Alice's outcome $A$ can be implemented using a quarter-wave plate and a polarizer before her measurement PBS. Alice's photons are randomly projected onto circular polarization states $\ket{R/L}{=}\frac{1}{\sqrt{2}}\left( \ket{H} {\pm} i \ket{V} \right)$---which are orthogonal to all measurements in the performed CHSH test---and re-prepared in eigenstates of the PBS to force one of the two outcomes $A=\pm1$. This approach breaks all relevant incoming causal arrows and allows deterministically setting the variable $A$, thus satisfying the conditions for an intervention. The measurement bases for Alice and Bob, as well as the setting of the intervention POL and QWP were chosen randomly using quantum random numbers from the Australian National University's online quantum random number generator based on Ref.~\cite{Symul2011}.

Single-photon clicks in the APDs for each outcome are registered with an AIT-TTM8000 time-tagging module with a temporal resolution of $82$~ps. Outcome probabilities, used to estimate $\mathrm{ACE}$, were computed from a total of 48,000 coincidence counts and no more than one event was registered for each set of random choices for $X,Y$, as well as the two elements of $I$.

\begin{figure}[t!]
  \begin{center}
\includegraphics[width=\columnwidth]{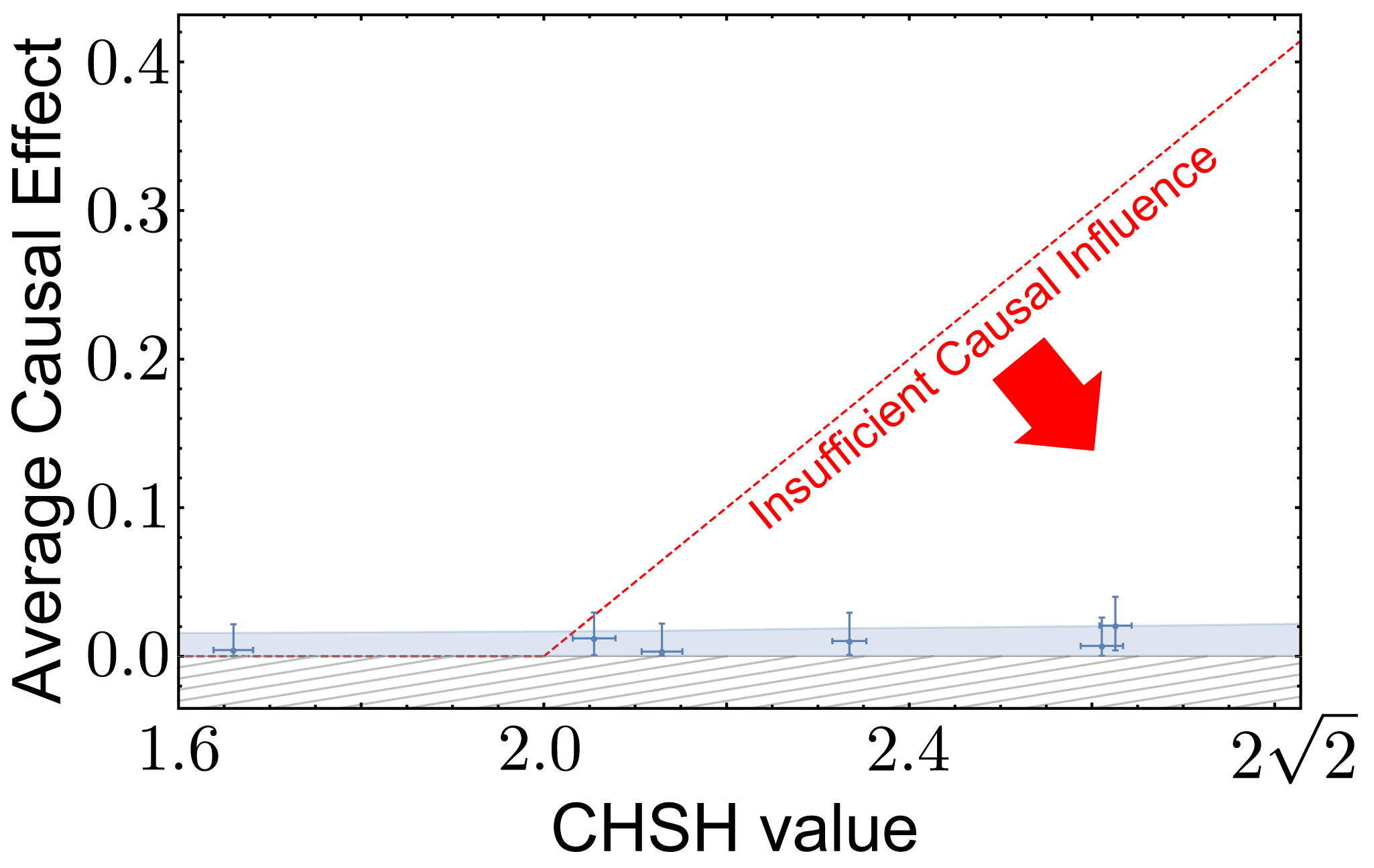}
  \end{center}
\vspace{-5mm}
\caption{\textbf{Observed average causal effect $\mathrm{ACE}$ versus measured CHSH-value.} Any value below the dashed red line, given by Eq.~\eqref{eq:ACE_CHSH}, is not sufficient to explain the observed CHSH-violation. Note, that the quantity $\mathrm{ACE}$ is bounded from below by 0, as indicated by the hatched area, resulting in asymmetric error distributions. The blue shaded area represent the 3$\sigma$ region of Poissonian noise. All errors represent $3\sigma$ statistical confidence intervals obtained from a Monte-Carlo simulation of the Poissonian counting statistics.}
  \label{fig:ResultsACE}
\end{figure}

Figure~\ref{fig:ResultsACE} shows the observed average causal effect as a function of the CHSH values measured for a range of entangled states. All measured values are below $\mathrm{ACE_{A\to B}}=0.02^{+0.02}_{-0.02}$ and largely independent of the observed CHSH violation. Note that the quantity is bounded from below, which results in non-Gaussian statistics and makes the value $0$ unachievable in the presence of experimental imperfections and finite counting statistics. When taking this in to account, all data lie within the 3$\sigma$ noise due to Poissonian counting statistics, see Sec~\ref{Sec:Supp4}. All quoted uncertainties were obtained from Monte Carlo simulations of the Poissonian counting statistics and correspond to the $0.13^\text{th}$ and $99.87^\text{th}$ percentile, respectively (in the case of normal distributed variables this would correspond to $3\sigma$ confidence regions). Within current experimental capabilities we find that CHSH violations above a value of $S_2=2.05\pm0.02$ cannot be fully explained by means of a direct causal influence from one outcome to the other. That is, the potential causal influence between Alice's and Bob's measurement---the green arrow in \figref{fig:models}b---is not sufficiently strong.

\textbf{Observational method ---}
As we have demonstrated, interventions are a powerful tool for quantifying the strength of causal influences, as measured by the average causal effect. Any experimental implementation of an intervention, however, relies on the quantum description of the degree of freedom responsible for the outcome $A$ (in the case above, the polarization), and is thus necessarily device dependent. We now show how moving beyond the CHSH scenario allows for a device independent test of any model with an arbitrarily strong causal influence from one outcome to the other.

Consider the situation where each of the two parties can choose to measure one of \emph{three} different dichotomic observables. As shown in Ref.~\cite{Chaves2015b}, any correlations compatible with the model in Fig.~\ref{fig:models}b must now satisfy
\begin{equation}
S_{3} = \langle E_{00} \rangle  - \langle E_{02} \rangle
- \langle E_{11} \rangle + \langle E_{12} \rangle
-\langle E_{20} \rangle + \langle E_{21} \rangle  \leq 4 .
\label{eq:I3322E}
\end{equation}
This inequality is symmetric and, as we show in Sec.~\ref{Sec:Supp1}, satisfied by any model that contains one-way communication of outcomes from either party to the other. Crucially, this allows us to test the models in Fig.~\ref{fig:models}b in a device-independent fashion and without committing to any particular temporal ordering of $A$ and $B$. 

\begin{figure}[h!]
  \begin{center}
\includegraphics[width=\columnwidth]{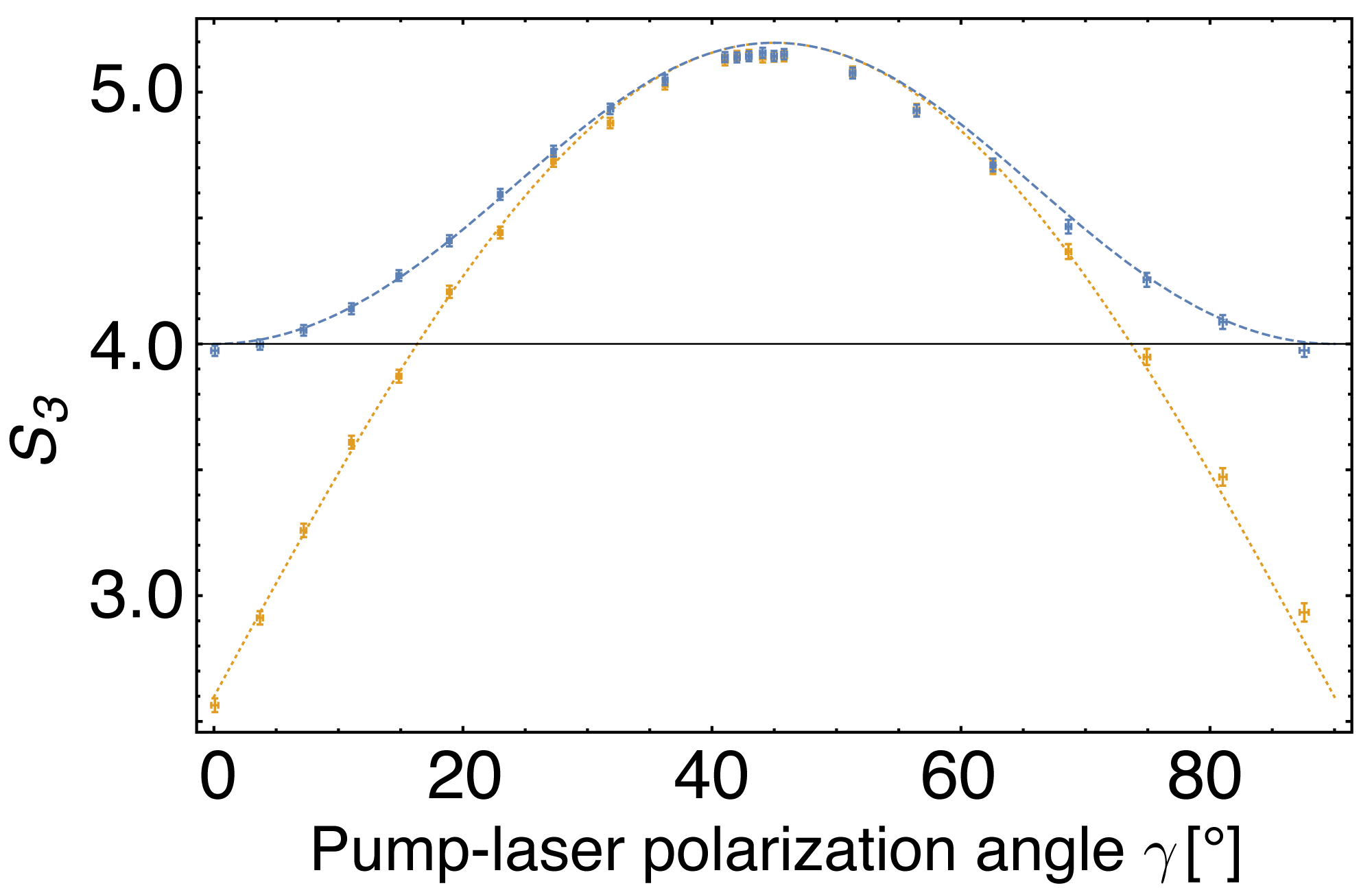}
  \end{center}
\vspace{-5mm}
\caption{\textbf{Observed values $S_3$ for a variety of quantum states of the form $\cos (\gamma) \ket{HV} + \sin (\gamma) \ket{VH}$.} The orange data points are observed using a fixed measurement scheme (optimal for the maximally entangled state, $\gamma=45^\circ$), with the dotted, orange line representing the corresponding theory prediction. The blue data and blue dashed theory line corresponds to the case where measurement settings were optimized for the prepared states, see Sec.~\ref{Sec:Supp3}. The black line represents the bound of inequality~\eqref{eq:I3322E}; any point above this line cannot be explained causally by a model of the form in Fig.~\ref{fig:models}b. Error-bars correspond to $3\sigma$ statistical confidence intervals.}
  \label{fig:Results3322}
\end{figure}

To test inequality~\eqref{eq:I3322E} Alice and Bob each perform measurements on their quantum system along one of three directions in the equatorial plane of the Bloch-sphere. These measurements are implemented using the setup in Fig.~\ref{fig:Setup} with the intervention elements $I$ removed. The specific measurement settings are given in Sec.~\ref{Sec:Supp3}. Figure~\ref{fig:Results3322} shows the observed violation of inequality~\eqref{eq:I3322E} as a function of the parameter $\gamma$ of the used quantum state. The theoretical maximal violation of the inequality is achieved using a maximally entangled state, corresponding to $\gamma=45^\circ$.

We observe a value of up to $S_{3}=5.16^{+0.02}_{-0.02}$, corresponding to a violation of Eq.~\eqref{eq:I3322E} by more than $170$ standard deviations. Complementary to the interventional experiment---which bounds the strength of a possible causal influence from one outcome to the other the CHSH scenario---this result rules out any causal model with an arbitrarily strong direct causal influence from one outcome to the other. This demonstrates that the conclusions of the interventional experiment hold even in a fully device-independent scenario and thus that a direct causal influence from one outcome to the other cannot explain quantum correlations.

\textbf{Discussion ---}
Previous work on causal explanations beyond local hidden variable models focused on testing Leggett's \emph{crypto nonlocality}~\cite{Groblacher2007,Paterek2007,Branciard2008}, a notion which concerns models with a very specific choice of hidden variable and that is in fact unrelated to Bell's local causality~\cite{Branciard2013}. In contrast, we make no assumptions on the form of the hidden variable and test all models compatible with the causal structure in Fig.~\ref{fig:models}b, which is a natural generalization of Bell's model and contains it as a special case. Practically, our experiment relies on a fair sampling assumption, see Sec.~\ref{Sec:Supp2}.

Our results demonstrate that causal modeling and interventions are powerful tools for studying causal explanations of quantum correlations beyond Bell's local hidden variable model. It would now be of considerable interest to further develop and extend these tools to test other classes of causal models, e.g.\ allowing for retrocausal influences or relaxations of measurement independence~\cite{Spekkens2015,Hall2010,Hall2011,Barrett2010,Chaves2015b,Gallicchio2014,Aktas2015,Putz2015}. 
Alternatively one could completely abandon the classical notion of causality and pursue a novel framework of quantum causality~\cite{Fritz2012,Fritz2014,cavalcanti2014modifications,pienaar2014graph,Chaves2015a,Leifer2013,Henson2015}. An important question for these approaches, however, is how to treat interventions and to what extent causal discovery remains possible \cite{costa2015quantum}. For example, interpretations of quantum mechanics that feature objective collapse of the wave-function would not permit interventions in the way they are used within the causal modelling framework~\cite{Naeger2015}.

Recent experiments put strong constraints on realist interpretations of quantum mechanics, ruling out maximally-epistemic~\cite{Ringbauer2015} and local-causal~\cite{hensen2015experimental,Shalm2015,Giustina2015} models. Our results exclude an important class of nonlocal causal models, thus contributing to a clearer picture of the status of reality and causality in quantum mechanics.


\begin{acknowledgments}
We thank C.\ Branciard, E.\ Cavalcanti and H.\ Wiseman for helpful discussions. We also thank the team from the Austrian Institute of Technology for kindly providing the time-tagging modules for this experiment. This work was supported in part by the Centres for Engineered Quantum Systems (CE110001013) and for Quantum Computation and Communication Technology (CE110001027), and the Templeton World Charity Foundation (TWCF 0064/AB38). RC acknowledges support from the Excellence Initiative of the German Federal and State Governments (Grants ZUK 43 \& 81), the US Army Research Office under contracts W911NF-14-1-0098 and W911NF-14-1-0133 (Quantum Characterization, Verification, and Validation), the DFG (GRO 4334 \& SPP 1798). AGW acknowledges support through a UQ Vice-Chancellor's Senior Research and Teaching Fellowship, and AF through an Australian Research Council Discovery Early Career Award, DE130100240.
\end{acknowledgments}

\onecolumngrid
\clearpage
\renewcommand{\theequation}{S\arabic{equation}}
\renewcommand{\thefigure}{S\arabic{figure}}
\renewcommand{\thetable}{\Roman{table}}
\renewcommand{\thesection}{S\Roman{section}}
\setcounter{equation}{0}
\setcounter{figure}{0}
\begin{center}
{\bf \large Supplementary Information \\
Experimental Test of Nonlocal Causality}
\end{center}
\medskip
\twocolumngrid
Here we discuss in detail the relation of the average causal effect to the CHSH violation. We also present the derivation of our novel 3-setting inequality and discuss data and error analysis.

\section{Relaxation of local causality}
\label{Sec:Supp1}
\noindent Here we will discuss in detail Bell's assumption of local causality and how it relates to assumptions on the underlying causal structure.

Local causality captures the idea that there should be no causal influence from one side of the experiment to the spacelike separated other side. Formally, this is a constraint on the conditional probability distribution: $p(a\vert b,x,y,\lambda)=p(a\vert x,\lambda)$. Here and in the following we will not explicitly state the equivalent constraint for Bob. We would like to stress that local causality is not equivalent to signal locality, which follows from special relativity and imposes constraints on the observable probabilities only: $p(a\vert x,y) = p(a\vert x)$. The natural generalization of signal locality to include the hidden variable is typically referred to as \emph{parameter independence} or  \emph{locality}: $p(a\vert x,y,\lambda) = p(a\vert x,\lambda)$~\cite{WisCav16}. Parameter independence together with what is often referred to \emph{outcome independence}: $p(a\vert b,x,y,\lambda) = p(a\vert x,y,\lambda)$ then implies local causality.

\begin{figure}[h!]
  \begin{center}
\includegraphics[width=\columnwidth]{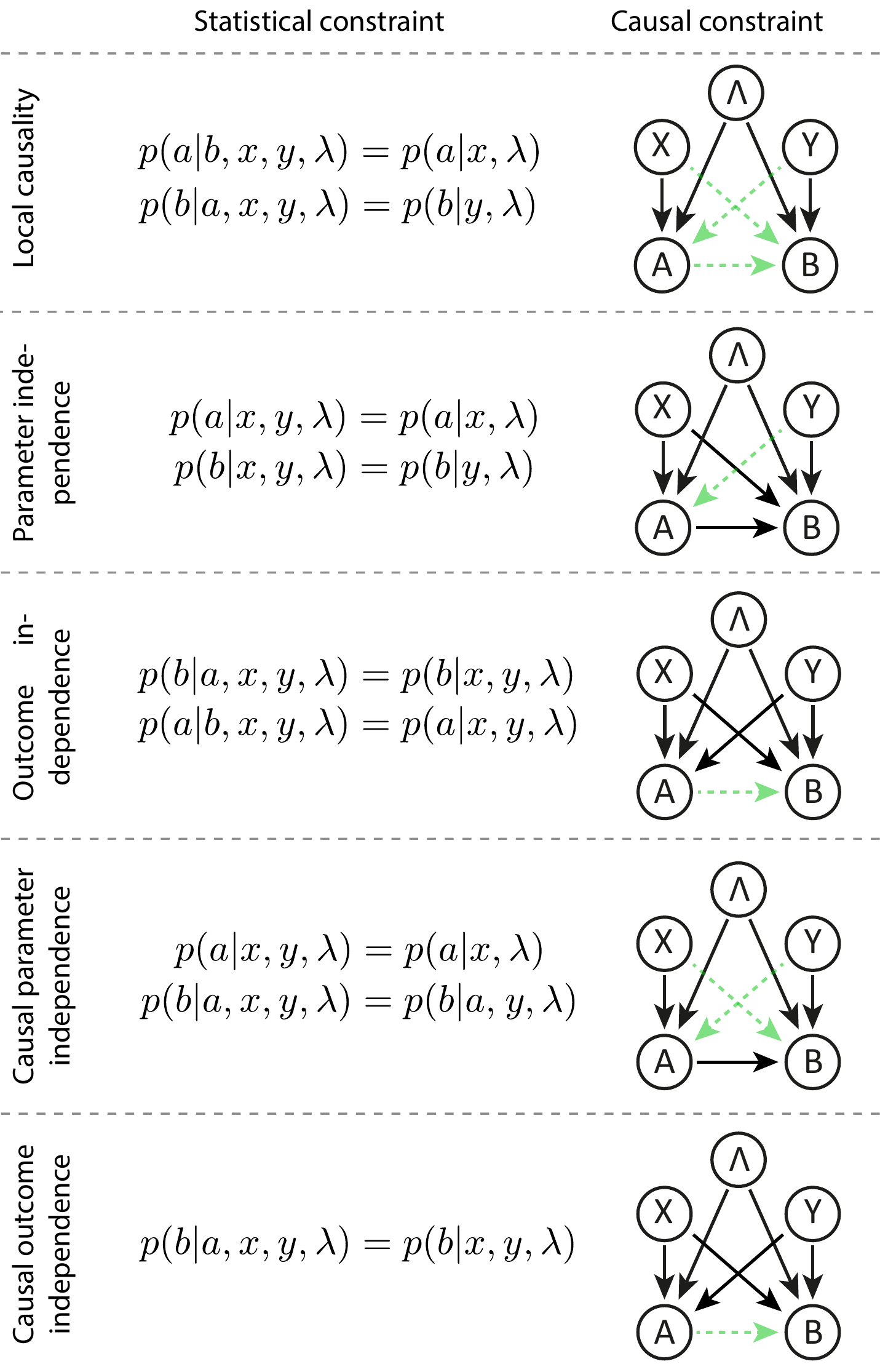}
  \end{center}
\vspace{-5mm}
\caption{\textbf{Comparision of various constraints on the causal structure of Bell's theorem.} The causal links forbidden by the respective assumption are shown in dashed green. Note that the statistical constraints implied by causal outcome independence and causal parameter independence are asymmetric in $a$ and $b$, and swapping them would result in a causal structure where the arrow between $A$ and $B$ is reversed. Our experimental test applies to both of these structures and any convex combination of them.}
  \label{fig:SuppAssumptions}
\end{figure}

Interpreted in the causal modelling framework local causality implies that there is no causal link from Bob's measurement setting $Y$ or outcome $B$ to Alice's measurement outcome $A$, cf. Fig.~\ref{fig:SuppAssumptions}. In the spirit of causal modelling we would like to obtain the causal structure of Bell's theorem directly from investigating these causal independencies. Specifically we denote by \emph{causal parameter independence} the absence of a causal link from Alice's measurement setting $X$ to Bob's measurement outcome $B$, and by \emph{causal outcome independence} the absence of a causal link from Alice's measurement outcome $A$ to Bob's measurement outcome $B$. 
As shown in Fig.~\ref{fig:SuppAssumptions}, causal outcome independence is indeed very similar to outcome independence. In contrast to the statistical variant of parameter independence, however, causal parameter independence indeed captures the idea that there should be no direct causal influence from the measurement setting on one side to the measurement outcome on the other side. In the following we will consider in detail the causal models that satisfy causal parameter independence, but may violate causal outcome independence. The corresponding probability distribution can be decomposed as,
\begin{equation}
\label{ACEmodel_app}
p(a,b\vert x,y)= \sum_{\lambda} p(a\vert x,\lambda) p(b\vert a,y,\lambda)p(\lambda).
\end{equation}

To understand how such a model allows for the generation of nonlocal correlations, consider the two following deterministic strategies:
\begin{eqnarray}
\label{strategies}
\text{strategy 1 } \rightarrow a=x, \text{ } b=a(y\oplus 1) \\ \nonumber
\text{strategy 2 } \rightarrow a=x \oplus 1, \text{ } b=y(a\oplus 1)\oplus a
\end{eqnarray}
Mixing the two strategies with equal probabilities generates the so-called Popescu-Rohrlich \cite{Popescu1994} distribution, $p(a,b \vert x,y)= (1/2)\delta_{a\oplus b,xy}$, which achieves maximal algebraic violation of the CHSH inequality, $S_2=4$. Crucially, however, despite the direct causal link from $A$ to $B$, the above example does not permit to send signals between Alice and Bob at the level of empirical observations, satisfying $p(b\vert x,y)=p(b\vert x^{\prime},y)$. In this case, however, signal locality does not follow from the causal structure in Fig.~1b), but rather from the precise choice of mixing probabilities in the above strategy. This kind of \emph{fine-tuning} of model parameters in order to ensure no-signalling has been found to be a common feature of all causal explanations of Bell correlations~\cite{Spekkens2015}.

The concept of fine-tuning plays an important role in causal discovery, which traditionally excludes fine-tuned models as unfaithful representations. The main justification for this step stems from the fact that assuming a uniform prior over the space of probabilistic parameters, the volume of it reproducing conditional independence relations not implied by the causal structure itself has measure zero ~\cite{Pearlbook,uhler2013geometry}. In practice, however, estimation error issues due to data of finite sample size can result in the volume of unfaithful parameters being considerably large \cite{uhler2013geometry}. From a purely causal inference perspective, this practical aspect---together with the fact that no faithful causal model can reproduce nonlocal correlations~\cite{Spekkens2015}---indicates that in order to conclusively eliminate fine-tuned models as a possible causal explanation to nonlocality, new methods are required.

As in the usual Bell scenario shown in Fig.~1a), each of the probabilities appearing in Eq. \eqref{ACEmodel_app} can be identified with a deterministic function. To see that, consider the general case where Alice has $m_x$ inputs and $o_a$ outputs, that is, $x=0,\dots,m_x-1$ and $a=0,\dots,o_a-1$ (and analogously for Bob).
The causal structure in Fig.~1b) assures that $a = f_A (x, \lambda)$ which resembles the usual LHV model and therefore implies $o_{a}^{m_x}$ different deterministic functions $f_A$. For $b$, however, we have that $b=f_B(a,y,\lambda)$ requiring $o_{b}^{o_a m_y}$ different deterministic functions $f_B$. That is, in order to fully describe the causal structure we need an underlying hidden variable with
$n=o_{a}^{m_x} o_{b}^{o_a m_y}$ possible values. In terms of these deterministic functions, the decomposition in \eqref{ACEmodel_app} can be rewritten as
\begin{equation}
\label{ACEmodel_ddet_app}
p(a,b\vert x,y)= \sum_{\lambda} \delta_{a,f_A(x,\lambda)} \delta_{b,f_B(a,y,\lambda)}p(\lambda).
\end{equation}

It is useful to represent $p(a,b|x,y)$ as a vector $\p$ with components $\p_j$ labeled by the multi-index $j=(a,b,x,y)$. Similarly, the distribution of $\Lambda$ can be represented by a vector with components $\q_\lambda=p(\Lambda = \lambda)$. It follows then that $\p = T \q$ where $T$ is a matrix with elements $T_{j,\lambda} = \delta_{a,f_A(x,\lambda)}
\delta_{b,f_B(a,y,\lambda)}$.

\subsection{Analytical derivation of the average causal effect $\min \mathrm{ACE}_{A \rightarrow B}$ in the CHSH scenario}
\noindent We now show analytically that the experimentally accessible \emph{average causal effect} is a suitable measure of causal influence in the CHSH scenario.

In general the direct causal effect
\begin{equation}
\mathcal{C}_{A \rightarrow B}= \sup_{b,y,a,a^{\prime}} \sum_{\lambda} p(\lambda) \vert p(b\vert do(a), y, \lambda)-p(b\vert do(a^{\prime}), y,\lambda )\vert ,
\label{eq:meas_causal}
\end{equation}
which quantifies the maximal shift (averaged over the unobservable variable $\Lambda$) in the probability of $B$ caused by interventions in $A$, can be used to quantify the strength of the causal link from $A$ to $B$. Indeed, it was shown in Ref.~\cite{Chaves2015b} that $\mathcal{C}_{A \rightarrow B}$ directly quantifies the degree of violation of causal outcome independence required for a causal explanation of the observed CHSH-violation as $\min \mathcal{C}_{A \rightarrow B}=\max \left[ 0,(S_2-2)/2 \right]$, where the maximum is taken over all eight symmetries of the CHSH quantity under relabelling of inputs, outputs, and parties~\cite{Clauser1969}. The quantity $\mathcal{C}_{A \rightarrow B}$, however, is not directly experimentally accessible, therefore precluding its use in an experimental test of the such a causal link.

Here we use an experimentally accessible variant of $\mathcal{C}_{A \rightarrow B}$, which does not require knowledge of the hidden variable, the \emph{average causal effect}.
\begin{eqnarray}
\label{eq:ACE_app}
\mathrm{ACE}_{A \rightarrow B} & & = \sup_{b,y,a,a^{\prime}}  \vert p(b\vert do(a), y)-p(b\vert do(a^{\prime}), y)\vert \\ \nonumber
 & & = \sup_{b,y,a,a^{\prime}}  \pm \left( p(b\vert do(a), y)-p(b\vert do(a^{\prime}), y) \right)
\end{eqnarray}
This expression quantifies the average causal effect from variable $A$ into variable $B$. That is, the minimum shift in the probability distribution of $B$ that we should observe by interventions on the variable $A$, if indeed the underlying causal structure is that shown in Fig.~\ref{fig:models}b. We are therefore interested in the following optimization problem
\begin{eqnarray}
\label{eq:opt_problem}
\underset{ {\bf q} \in \RR^{n}}{\minimize} & & \quad \mathrm{ACE}_{A \rightarrow B}  \\ \nonumber
\st & & \quad  T{\bf q} =  \p  \\ \nonumber
& & \quad \langle \1_n, {\bf q} \rangle = 1  \\ \nonumber
& & \quad {\bf q}  \geq \0_n.
\end{eqnarray}
The two last constraints following from the fact that the hidden variable $\Lambda$ should be described by a well defined probability distribution (positive and normalized).

To write the optimization problem Eq.\eqref{eq:opt_problem} as a standard linear program, notice that
\begin{align}
&  \left( p(b\vert do(a), y)-p(b\vert do(a^{\prime}), y) \right)\\ \label{eq:aux}
&= \sum_{\lambda}  p(\lambda) \left( \delta_{b,f_B(a,y,\lambda)} - \delta_{b,f_B(a^{\prime},y,\lambda)}\right) \\
&= \sum_{i} q_i v_i= \langle \vv, \q \rangle,
\end{align}
where the vector ${\bf v}={\bf v}(a,a^{\prime},y,b)$ fully characterizes the action of the Kronecker-symbols in eq. (\ref{eq:aux}).
The ACE measure \eqref{eq:ACE_app} can then be recast as
\begin{equation}
\label{ACE_2_app}
\mathrm{ACE}_{A \rightarrow B} = \max_{i=1,\dots,2L} \langle {\bf q} , {\bf v}_i \rangle = C {\bf q}.
\end{equation}
Here, the index $i$ parametrizes the $2L$ possible instances of $(a,a^{\prime},y,b)$ with $x \neq x^{\prime}$ (the factor $2$ coming from the $\pm$ signs in Eq. \eqref{eq:ACE_app}) and $\vv_i = v(a,a',y,b)$ denotes the vector corresponding to that instance. The matrix $C$ subsumes all these different instances, that is, $C := \sum_{i=1}^L | \e_i \rangle \langle \vv_i |$ where $\e_i$ stands for an orthonormal basis.

The optimization problem in Eq.\eqref{eq:opt_problem} is then equivalent to a standard linear program
\begin{eqnarray}
\label{eq:opt_problem2}
\underset{ {\bf v,q}}{\minimize} & & \quad v  \\ \nonumber
\st & & \quad  T{\bf q} =  \p  \\ \nonumber
& & \quad \langle \1_n, {\bf q} \rangle = 1  \\ \nonumber
& & \quad {\bf q}  \geq \0_n \\ \nonumber
& & \quad C {\bf q} \leq v.
\end{eqnarray}

As proved in Ref.~\cite{Chaves2015b}, to obtain the solution of this problem for any vector $\p$ encoding the full probability distribution $p(a,b \vert x,y)$, we have to consider the dual optimization problem. In practice, that means that solving Eq.~\eqref{eq:opt_problem2} is equivalent to evaluating
\begin{equation*}
\max_{1 \leq i \leq K} \langle \d_i, \p \rangle,
\end{equation*}
where $\left\{ {\bf d}_i \right\}_{i=1}^K$ denotes the vertices of the dual feasible region (see Ref.~\cite{Chaves2015b} for further details).

We have performed such an analysis for the particular case of the CHSH scenario ($m_x=m_y=o_a=o_b=2$) using PORTA~\cite{porta}, a standard software for the evaluation of extremal points of a polyhedron. Similarly to what happens to the measure $\mathcal{C}_{A \rightarrow B}$ (see Suplemental material of Ref.~\cite{Chaves2015b}), the extremal points of the dual region correspond to non-signalling constraints and all the symmetries of the CHSH inequality (up to a constant factor), implying that
\begin{equation}
\min \mathrm{ACE}_{A \rightarrow B}=\max \left[ 0,(S-2)/2 \right].
\label{eq:ACE_CHSH_app}
\end{equation}

\subsection{Proving the new inequality}
\noindent As shown above, a relaxation of causal outcome independence, while maintaining causal parameter independence, allows for the classical explanation of any nonlocal correlations in the CHSH scenario, where the two parties perform two possible dichotomic measurements. However, if the parties perform three or more measurements each, this does not hold anymore.

The decomposition \eqref{ACEmodel_ddet_app} defines a polytope of correlations that are compatible with the causal model in Fig.~\ref{fig:models}b. This polytope is characterized by $n=o_{a}^{m_x} o_{b}^{o_a m_y}$ extremal points. Therefore, for a fixed number of measurements and outcomes, one can resort to usual convex optimization software in order to find its description in terms of finitely many Bell inequalities. As shown in Ref.~\cite{Chaves2015b}, one of the Bell inequalities characterizing the polytope in the case $o_{a}=o_{b}=2$ and $m_x=m_y=3$ is given by inequality~\eqref{eq:I3322E}. An easy way of proving that this inequality in fact holds, is to verify that for each of the $n=2^32^6$ extremal points defining the polytope this inequality is satisfied.

A similar argument can be used to prove that this inequality is also valid if we reverse the roles of Alice and Bob. That is, in this case we allow the outcomes of Alice to depend on the outcomes of Bob:
\begin{equation}
\label{ACEmodel_app_new}
p(a,b\vert x,y)= \sum_{\lambda} p(a\vert x,b,\lambda) p(b\vert y,\lambda)p(\lambda).
\end{equation}
Since both the models Eq.~\eqref{ACEmodel_app} and Eq.~\eqref{ACEmodel_app_new} respect inequality~\eqref{eq:I3322E}, so does a convex combination of both of them. In other words, any model of the form
\begin{eqnarray}
\label{eq:out_ind}
p(a,b \vert x,y)  = & & \sum_{\lambda} p(a\vert x, \lambda) p(b\vert y,a, \lambda) p(\lambda)+ \\ \nonumber
 & & \sum_{\mu} p(a\vert x,b, \mu) p(b\vert y, \mu) p(\mu),
\end{eqnarray}
with $\sum_{\mu,\lambda} p(\mu)+p(\lambda)=1$, respects inequality~(5). Any pure two-qubit entangled state, however, can generate correlations violating this inequality~\cite{Chaves2015b}. This allows us to show unambiguously and based on observational data only, that a direct causal influence from one outcome to the other cannot explain quantum correlations.\\

\section{Theoretical analysis of experimental imperfections}
\label{Sec:Supp2}
\noindent Any experiment suffers from imperfections in the form of detector inefficiencies, noise and other possible forms of loss. Practically our experiment thus relies on a fair-sampling assumption. Here we provide a short analysis of the requirements for testing inequality~\eqref{eq:I3322E} without this assumption. Our analysis is similar to what is usually done for Bell inequalities, for example, the CHSH inequality~\cite{Eberhard1993,Percival2008}

To describe the inefficiency of the photon detectors, we model the projective measurements being performed in the experiment via the following POVM with elements~\cite{ChavesBrask2011}
\begin{equation}
\label{povm}
\begin{split}
M_\uparrow & = \eta_\uparrow \ket{\uparrow}_s\bra{\uparrow} + (1-\eta_\downarrow) \ket{\downarrow}_s\bra{\downarrow} , \\
M_\downarrow & = \eta_\downarrow \ket{\downarrow}_s\bra{\downarrow} +
(1-\eta_\uparrow)\ket{\uparrow}_s\bra{\uparrow}.
\end{split}
\end{equation}
For perfect efficiencies $\eta_{\downarrow,\uparrow} = 1$ this POVM implements a projective measurement along the direction defined by $s$, that is, $\ket{\uparrow}_s,\ket{\downarrow}_s$ are the eigenstates of the observable $\ket{\uparrow}_s\bra{\uparrow}-\ket{\downarrow}_s\bra{\downarrow}$. Following the usual treatment of detection inefficiencies~\cite{Eberhard1993,Percival2008,ChavesBrask2011}, given a non-click event of the detectors happening with probability $\eta$ (assumed to be the same for both detectors), we bin it together with the $\downarrow$-outcome. That is, our faulty measurements are described by Eq.~\eqref{povm} with $\eta_\downarrow=1$ and $\eta_\uparrow=\eta$.

We also consider errors in the preparation of the states. To that aim we consider the initial two-qubit state to be affected by white noise (parameterized by the visibility $v$):
\begin{equation}
\rho_{\gamma}(v)=v\ket{\Psi_{\gamma}}\bra{\Psi_{\gamma}}+(1-v)\mathbb{I}/4
\end{equation}
where $\Psi_{\gamma}=\cos \gamma \ket{HV} + \sin \gamma \ket{VH}$ represents the pure two-qubit entangled state that we would ideally prepare.

\begin{figure}[h!]
  \begin{center}
\includegraphics[width=\columnwidth]{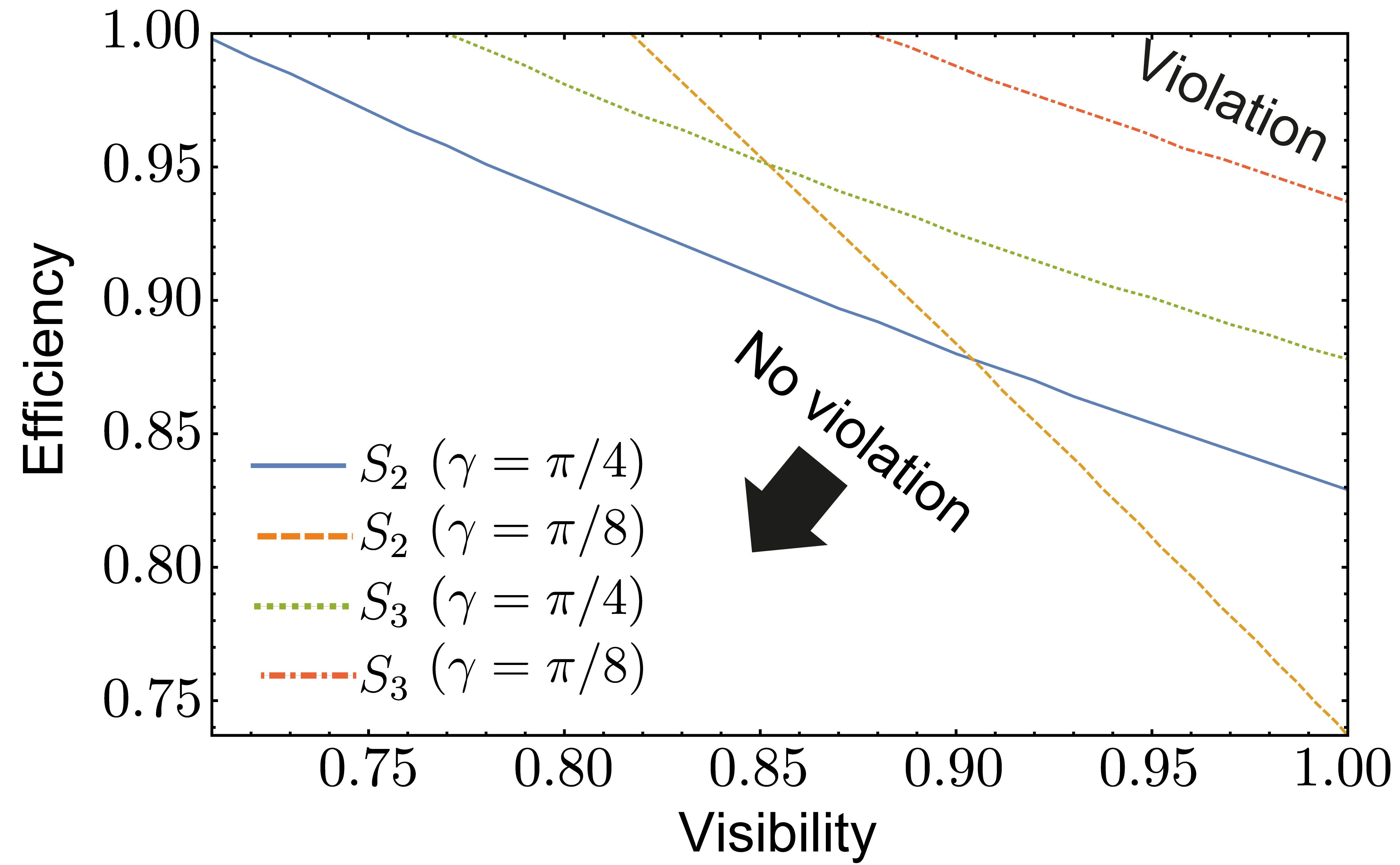}
  \end{center}
\vspace{-5mm}
\caption{\textbf{Efficiency $\eta$ and visibility $v$ requirements for a violation of inequality~\eqref{eq:I3322E} without fair-sampling assumption.} For a maximally entangled state ($\gamma=\pi/4$) inequality~\eqref{eq:I3322E} is violated above the green, dashed line, while the CHSH inequality is violated above the blue, solid line. For a non-maximally entangled state ($\gamma=\pi/8$) the corresponding lines are red dot-dashed for inequality~\eqref{eq:I3322E} and orange, dashed for CHSH.}
  \label{fig:efficiency}
\end{figure}

For different values of $\gamma$ we have performed a numerical optimization over all possible POVMs Eq.~\eqref{povm} in order to find the minimum values of $v$ and $\eta$ leading to the violation of inequality~(5) without requiring the fair-sampling assumption. The results are shown in Fig.~\ref{fig:efficiency}. As can be seen, the smaller is the initial entanglement (parameterized by the angle $\theta$) the less robust is the violation of the inequality as function of the visibility and detector inefficiency. For a maximally entangled state the minimum required visibility and efficiency are, respectively, $v_{\mathrm{crit}} \sim 0.77$ and $\eta_{\mathrm{crit}}\sim 0.88$ For comparison, we also compute the requirements for the usual CHSH inequality~\cite{Eberhard1993,Percival2008}. Not surprisingly, since our inequality excludes a larger class of causal models, it has more stringent experimental requirements.

\section{Testing the new inequality}
\label{Sec:Supp3}
\noindent Similarly to the standard CHSH inequality, our 3-setting inequality~\eqref{eq:I3322E} can be tested using only measurements in the equatorial plane of the form $O=\cos(\theta)\hat Z+ \sin(\theta)\hat X$. Numerical results suggest that this is indeed optimal and the maximal violation of the inequality is achieved for $\theta^{(A)}_0= - \pi/6$, $\theta^{(A)}_1=7\pi/6$, $\theta^{(A)}_2=\pi/2$, $\theta^{(B)}_0= - \pi/3$, $\theta^{(B)}_1 = \pi/3$ and $\theta^{(B)}_2 = \pi$, where the superscript denotes the party and the subscript denotes the number of the measurement. Using these measurement settings we obtain
\begin{equation*}
S_{3}=\frac{3}{2} \sqrt{3} (1+ \sin (2 \gamma)).
\end{equation*}
Note that since $\mathcal{C}=|\sin 2 \gamma|$ this corresponds to a linear relationship between $S_{3}$ and the concurrence of the used state.

\begin{figure}[h!]
  \begin{center}
\includegraphics[width=\columnwidth]{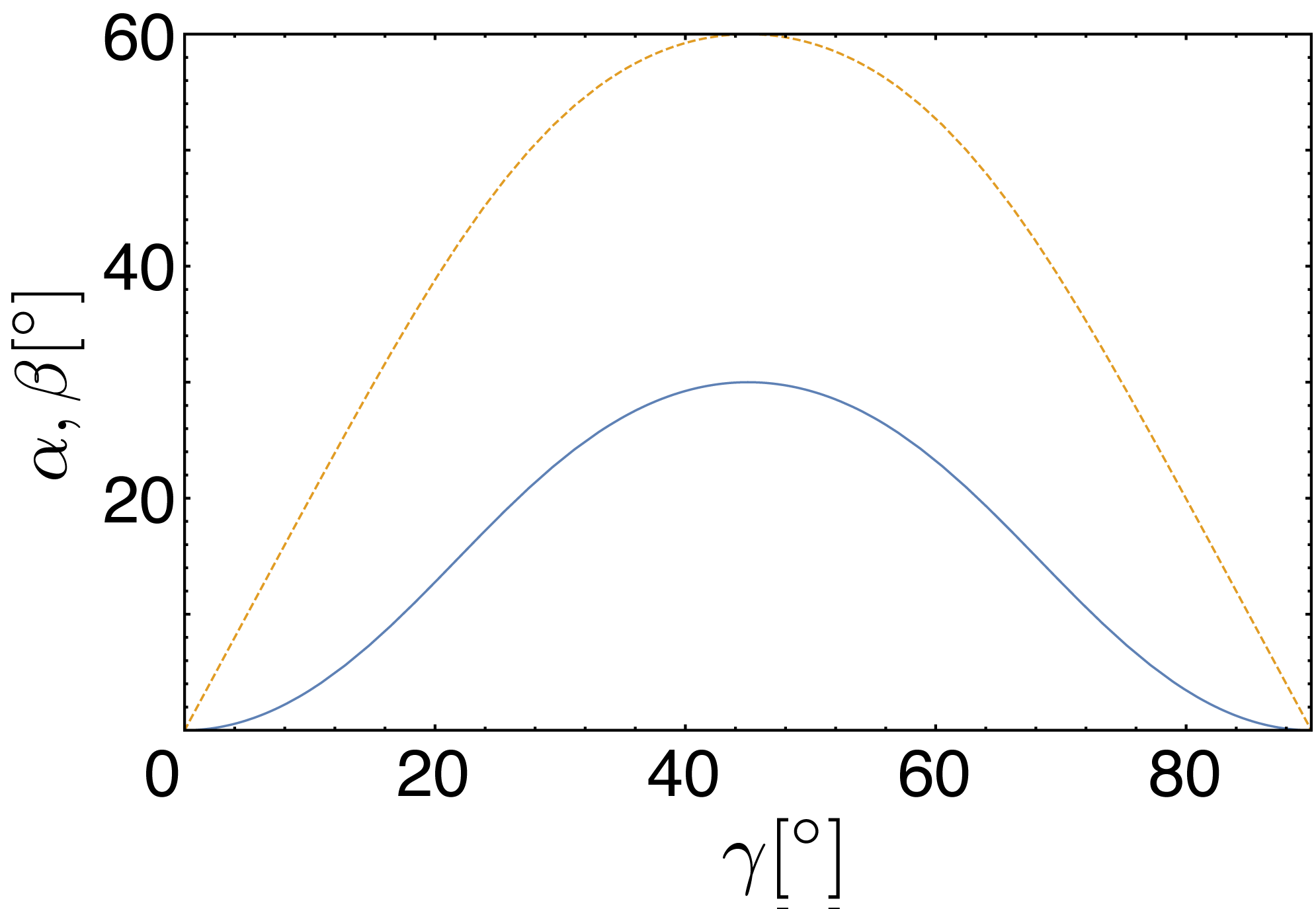}
  \end{center}
\vspace{-5mm}
\caption{\textbf{Measurement angles for inequality~\eqref{eq:I3322E}} Alice's measurement angle $\alpha$ shown in solid, blue, and Bob's measurement angle $\beta$ in dashed, orange.}
  \label{fig:SI:angles}
\end{figure}

As pointed out earlier, inequality~\eqref{eq:I3322E} can indeed be violated by every entangled quantum state. This is clearly demonstrated by $\theta^{(A)}_0= - \alpha$, $\theta^{(A)}_1=\alpha+\pi$, $\theta^{(A)}_2=\frac{\pi}{2}$, $\theta^{(B)}_0= - \beta$, $\theta^{(B)}_1 = \beta$ and $\theta^{(B)}_2 = \pi$, where the optimal angles $\alpha,\beta$ can be found analytically as a function of the state parameter $\gamma$. In this case we obtain:
\begin{align*}
S_{3}=&\left(\frac{\sqrt{\cos (4 \gamma)+7}-\sqrt{2} (\cos (2 \gamma)-3)}{2 \cos ^2(\gamma)} \right) \nonumber \\
&\quad \times \;\left( \sqrt{\cos (2 \gamma)+\sqrt{2} \sqrt{\cos (4 \gamma)+7}-3}\right) 
\end{align*}
The corresponding angles $\alpha$ and $\beta$ are shown in Fig.~\ref{fig:SI:angles}.

\section{Error analysis}
\label{Sec:Supp4}
\noindent Here we discuss the details of the error analysis for both experiments above.

\begin{figure}[h!]
  \begin{center}
\includegraphics[width=\columnwidth]{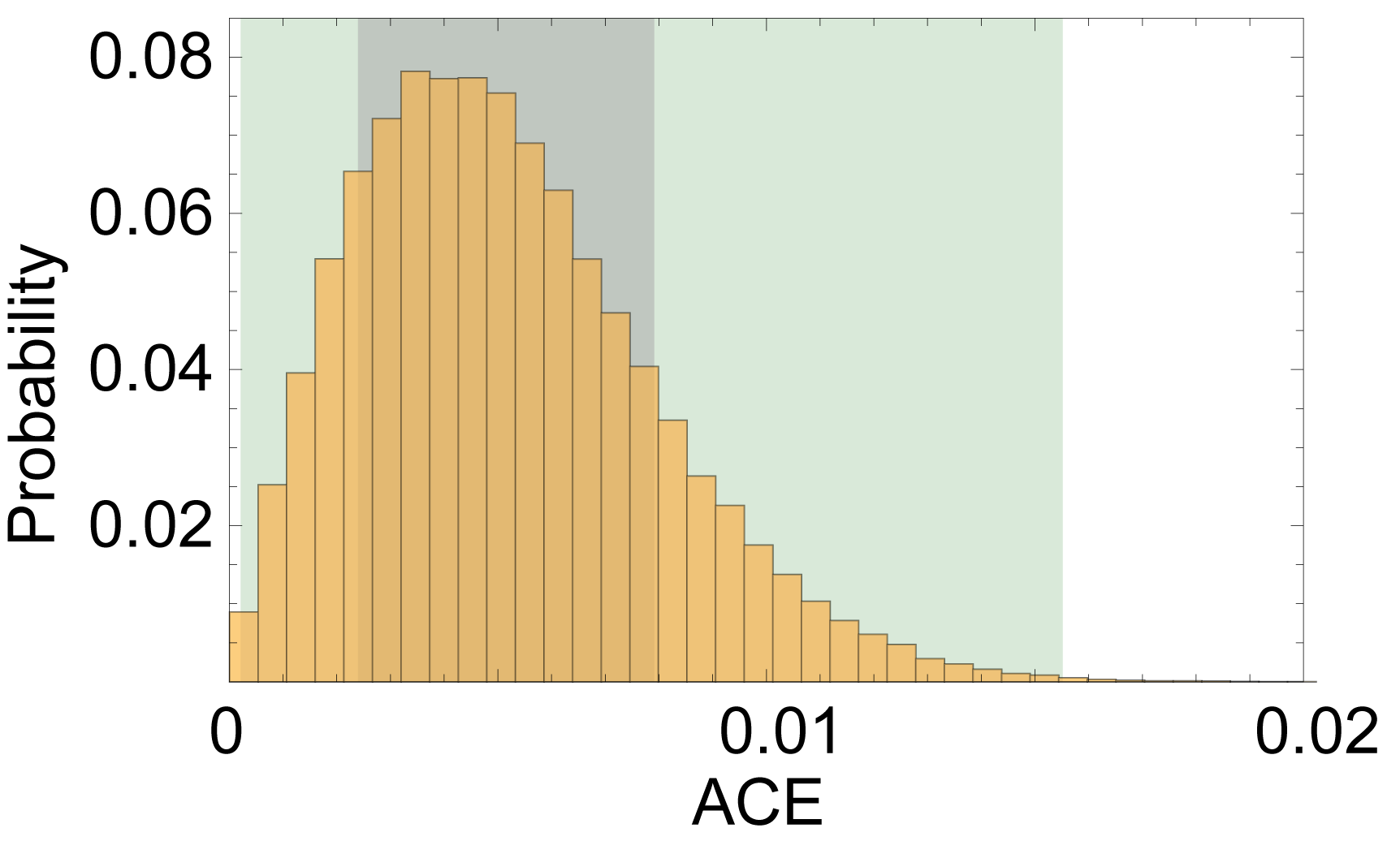}
  \end{center}
\vspace{-5mm}
\caption{\textbf{Distribution of statistical noise due to Poissonian counting statistics}. The dark, purple shaded area corresponds to the $68.27\%$ ($1\sigma$) confidence interval, while the light, green area is the $99.73\%$ ($3\sigma$) interval. These have been chosen for comparability with normal distributed data. Data was obtained from 100,000 runs of a Monte-Carlo simulation of the Poissonian counting statistics for perfect measurements.}
  \label{fig:SI:ACEerrorDist}
\end{figure}

Any photonic experiment suffers from statistical noise due to the Poissonian nature of the single-photon source and detection. Additionally it is a common feature for experiments measuring bounded quantities (such as $\mathrm{ACE}$) that observed distributions feature significant asymmetry close to the boundaries, making the observation of extreme values very unlikely. To illustrate this effect, Fig.~\ref{fig:SI:ACEerrorDist} shows simulated statistics for $\mathrm{ACE}$ assuming perfect measurement and only Poissonian noise at a total number of 48000 single-photon events. For our choice of quantum states, the distribution of single-photon clicks is such, that higher entangled states are more susceptible to this form of noise. As a consequence, the median of the distribution increases with entanglement from $0.0048$ to $0.0068$, with $3\sigma$ confidence intervals of $[0.0002, 0.0155]$ and $[0.0003, 0.0219]$, respectively. Note that all $3\sigma$-intervals quoted in this manuscript correspond to the intervals that contain $\sim 99.73\%$ of the data (in analogy with the $3\sigma$-region for a normal distribution) and are thus asymmetric around the median of the distribution, see Fig.~.\ref{fig:SI:ACEerrorDist}.

Besides the statistical errors there are various sources of systematic errors, which explain the consistent offset of the measured $\mathrm{ACE}$ from 0. For the interventionist experiment it is crucial that all CHSH measurements are performed in the equatorial plane, while the intervention acts on the poles of the Bloch sphere. The relevant waveplates are one HWP each to set Alice' and Bob's measurement basis, and one QWP for the intervention. The relative phase-shifts imparted by these waveplates and their accuracy are listed in Tab.~\ref{tab:Appendix:WPs}. These errors result in a tilt of the intervention from the orthogonal orientation of $0.0109^{+0.0122}_{-0.0009}$ and Alice's and Bob's measurement planes are tilted by $0.1403^{+0.0004}_{-0.0005}$ and $0.050^{+0.013}_{-0.010}$, respectively.

\begin{table}[h!]
\begin{tabular}{c@{\hskip 2em} c@{\hskip 2em} c}
\hline\hline
Element & $\Delta\phi (^\circ)$ & $r$ \\
\hline
HWP (X) & $0.04$ & $0.955$ \\
HWP (Y) & $0.04$ & $0.978$ \\
QWP (I) & $0.13$ & $0.503$ \\
\hline\hline
\end{tabular}
\caption{Shown are the standard deviations $\Delta\phi$ in the angles of the optical axes as determined from fits to the measured coincidence counts, as well as the retardance $r$ (in measures of $\lambda/2$) obtained from the visibility of the fringes observed in the same data, for the relevant waveplates.}
\label{tab:Appendix:WPs}
\end{table}
The accuracy of the intervention polarizer is $\Delta\phi=0.14^\circ$, with a contrast better than 7000:1. Alice's and Bob's measurement PBS have a contrast of greater than $500:1$ and $8000:1$, respectively and are aligned to each other to a contrast of $1000:1$. All of these systematic and statistical errors were taken into account in a Monte Carlo simulation. We observe only a slight increase in the median values and variance of the expected uncertainty distributions, and the systematic imperfections result in an additional offset from 0. This suggests that the offset in the data of Fig.~3 could be explained by systematic imperfections.

\bibliography{CausalModels}
\bibliographystyle{apsrev}

\end{document}